\documentclass[12pt]{elsarticle}
\usepackage{graphicx}
\usepackage{epsfig}
\usepackage{amsmath}
\usepackage{amssymb}
\usepackage{amsfonts}
\usepackage{color}
\journal{Physica A 392,1788 (2013)}

\begin{document}
\begin{frontmatter}
\title{Study of a Market Model with Conservative Exchanges on Complex Networks}

\author[ump,cps]{Lidia A.  Braunstein}
\ead{lbrauns@mdp.edu.ar}
\author[ump]{Pablo A. Macri}
\ead{macri@mdp.edu.ar}
\author[if,ppge]{J. R. Iglesias}
\ead{roberto@if.ufrgs.br}

\address[ump]{IFIMAR Institute, Physics Department, UNMdP-CONICET, Mar del Plata, Argentina}
\address[cps] {Center for Polymer Studies, Boston University, Boston,
  Massachusetts 02215, USA}
\address[if]{Instituto de F\'{\i}sica, UFRGS and Instituto Nacional de
Ci\^encia e Tecnologia de Sistemas Complexos, Caixa Postal 15051, 91501-970
Porto Alegre, RS, Brazil}
\address[ppge]{Programa de P\'os-Gradua\c{c}\~ao em Economia, UFRGS, Av. Jo\~ao Pessoa 52, 90040-000
Porto Alegre, RS, Brazil}
\date{\today}

\begin{abstract}
Many models of market dynamics make use of the idea of conservative wealth exchanges among economic agents.
A few years ago an exchange model using extremal dynamics was developed and a very interesting result was obtained: a self-generated minimum wealth or poverty line. On the other hand, the wealth distribution exhibited an exponential shape as a function of the square of the wealth. These results have been obtained both considering exchanges between nearest neighbors or in a mean field scheme. In the present paper we study the effect of distributing the agents on a complex network. We have considered  archetypical complex networks: Erd\"{o}s-R\'enyi random networks and scale-free networks. The presence of  a poverty line with finite wealth is preserved but spatial correlations are important, particularly between the degree of the node and the wealth. We present a detailed study of the correlations, as well as the changes in the Gini coefficient, that measures the inequality, as a function of the type and average degree of the considered networks.
\end{abstract}
\end{frontmatter}

\section{Introduction}

An empirical study focusing the income distribution of workers, companies and countries was first presented more than a century ago by Italian economist Vilfredo Pareto. He observed, in his book ``Cours d'Economie Politique''~\cite{Pareto}, that the distribution of income does not follow a Gaussian distribution but a power law. That means that the asymptotic behavior of the distribution function follows a power function that decreases, for big values of the wealth, as $w^{-\alpha}$, being $\alpha > 1$ the exponent of the power law. Non-gaussian distributions are denominated L\'evy distributions~\cite{Mantegna}, thus this power law distribution is nowadays known as Pareto-L\'evy Distribution. The exponent $\alpha$ is named Pareto index. The value of this exponent changes with geography and time, but typical values are close to $3/2$. The bigger the value of the Pareto exponent the higher the inequality in a society.

More recent wealth distribution statistics indicate that, even though Pareto distribution provides a good fit in the high income range, it does not agree with the observed data over the middle and low income ranges. For instance, data from Japan~\cite{souma}, Italy~\cite{clementi}, India~\cite{sinha1}, the United States of America and the United Kingdom~\cite{dragu2000,dragu2001a,dragu2001b} are fitted by a log-normal or Gaussian distribution with the maximum located at the middle income region plus a power law for the high income strata.

Power laws are not exceptions in nature~\cite{PerBak}, so it is not surprising that the wealth distribution follows a power law. The quiz with the income and wealth distribution is not the power law, but how this distribution is generated through the dynamics of the agents interacting. On the other hand, a Pareto-L\'evy distribution is more unequal than a Gaussian: when the distribution follows a power law there are more affluent agents than in the case of a Gaussian distribution, but also more impoverished agents.

In order to try to describe the processes that generate a given profile for the wealth distribution diverse exchange models have been widely applied to describe wealth and/or income distributions in social systems. Different mathematical models of capital exchange among economic agents have been proposed trying to explain these empirical data (for a review see ref.~\cite{Caon2007}). Most of these models consider an ensemble of interacting economic agents that exchange a fixed or random amount of a quantity called ``wealth''. This wealth represents the agents welfare. The exact choice of this quantity is not straightforward, but one can think that it stands for the exchange of a given amount of money against some service or commodity. Within these models the amount of exchanged wealth when two agents interact corresponds to some economic ``energy'' that may be randomly exchanged. If this exchanged amount corresponds to a random  fraction of one of the interacting agents wealth, the resulting wealth distribution is, unsurprisingly, a Gibbs exponential distribution~\cite{dragu2000} that fits the wealth distribution for the low and middle income range.

Aiming at obtaining distributions with power law tails, several methods
have been proposed. Numerical procedures \cite{Caon2007,chatter1,chatter2,chakra,sinha2,IGVA2004}, as well as some analytical calculations~\cite{Bouchaud,cristian07}, indicate that one frequent result of that kind of models is condensation, i.e. concentration of all available wealth in just one or a few agents~\cite{rita2012}. This result corresponds to a kind of equipartition of poverty: all agents (except for a set of zero measure) possess zero wealth while few ones concentrate all the resources. In any case, an almost ordered state is obtained, and this is a state of equilibrium, since agents with zero wealth cannot participate in further exchanges. The Gini coefficient~\cite{Gini} of this state is equal to 1, indicating perfect inequality~\cite{Caon2007,cristian07,rita2012}. Several methods have been proposed to avoid this situation, for instance, exchange rules where the poorer are favored~\cite{sinha1,Caon2007,IGVA2004,cristian07,west} but in all circumstances the final state is one with high inequality, i.e. very near condensation.

A few years ago one of us presented an alternative model for wealth distribution, the Conservative Exchange Market Model (CEMM), inspired by the ideas of John Rawls~\cite{Rawls} and Amartya Sen~\cite{Sen} and also by the Bak-Sneppen model for extinction of species~\cite{BakSnep}. The main point of the model is that some kind of action should be taken to change the state of the poorest agent in the society. The idea of a society that takes measures in order to improve the situation of the most impoverished is compatible with the propositions of John Rawls, in his book ``A Theory of Justice''~\cite{Rawls}, directed towards an inventive way of securing equity of opportunities as one of the basic principles of justice. He asserts that {\it no redistribution of resources within a state can occur unless it benefits the least well-off: and this should be the only way to prevent the stronger (or richer) from overpowering the weaker (or poorer)}. The practical way to carry out this proposition in a simulation was adapted from the Bak-Sneppen model for extinction of biological species~\cite{BakSnep}. In this model the less fitted species disappears and is replaced by a new one with different fitness, then, the appearance of this new species affects the environment changing the fitness of the neighboring species. In 2003 a similar model was developed where the role of the fitness is substituted by the assets of a particular agent, and the model is now conservative, the difference between the new and the old assets of the minimum wealth agent is taken from (if positive) or given to (if negative) the neighbors of the poorest agent~\cite{PIAV2003,SI2004}. The distribution obtained follows an exponential law as a function of the square of the wealth and a poverty line with finite wealth is obtained by self-organization, i.e. the poorer agents posses finite endowments (different to what happens in most exchange models where the minimum wealth is zero). The model was also studied in a kind of mean-field version where the poorest agent interact with randomly chosen agents~\cite{IGPVA2003,IGVA2004} or by selecting at random as partner of the agent with the minimal value of wealth one of the neighbors of the first agent, and then both agents re-shuffle their entire amount of wealth~\cite{manna}. Finally, redistribution with the full society was also considered~\cite{Igles2010}. The obtained Gini coefficient is relatively low and compares well with the values of the Gini coefficient of some Northern European countries as Denmark or Sweden~\cite{SI2004}. This suggests a path to decrease inequality in real societies~\cite{SI2004,Igles2010}. A similar model was presented recently by Ghosh {\it et. al.}~\cite{Ghosh} where particles below a given arbitrary threshold may interact with other particles. They studied the system in mean-field, $1D$ and $2D$ and found a distribution that deviates from the exponential Boltzmann-Gibbs one. Also, a phase transition in the number of particles below the poverty line is found as a function of the value of the threshold. This phase transition is not present in the CEMM theory because here the threshold is a result of the self-organization induced by the extremal dynamics.

Here we revisit this model considering a more realistic description of social networks. We study the system on scale-free  network (SF) and on  random Erd\"{o}s-Renyi (ER) network (for a review on these networks see, for example, ref.~\cite{AB}). The case of Watts-Strogatz small world networks has already been partially discussed in ref.~\cite{IGVA2004} and the results do not differ in a significative way from the mean field results of ref .~\cite{PIAV2003}. Besides, we do not try to describe the power law region of the wealth distribution, as we are only considering additive exchanges. The results will describe the middle and low income agents, and also, we will obtain a poverty line that is a characteristic of some countries, like Scandinavian ones where protective measures are taken to help the less favored people. Anticipating on the results we are going to show that the poverty line is a robust characteristic of the model, but a strong correlation between the connectivity of the agents and their wealth is also observed. Indeed, it seems that the more connected agents exhibit a wealth less than average, because they have higher probability of interacting with poorer agents.

In the next section we present a very short review of the original model: the Conservative Exchange Market Model (CEMM) and its main conclusions. Then, in section III we introduce the SF and ER networks and present the main results for these networks. In section IV we discuss the results and present the conclusions.

\section{The Conservative Exchange Market Model - CEMM}
\label{CEMM}

The Conservative Exchange Market Model (CEMM)~\cite{PIAV2003} is a simple macroeconomic model that consists of a one-dimensional lattice with $N$ sites and periodic boundary conditions. Each site represents an economic agent (individuals, industries or
countries) linked to two neighbors. To each agent it is assigned some wealth-parameter that represents its welfare, like, for example, the GDP for countries or accumulated wealth for individuals. One chooses an arbitrary initial configuration where the wealth is a number between 0 and 1 distributed randomly and uniformly among agents.

\begin{figure}
\centering
\includegraphics[width=8.cm]{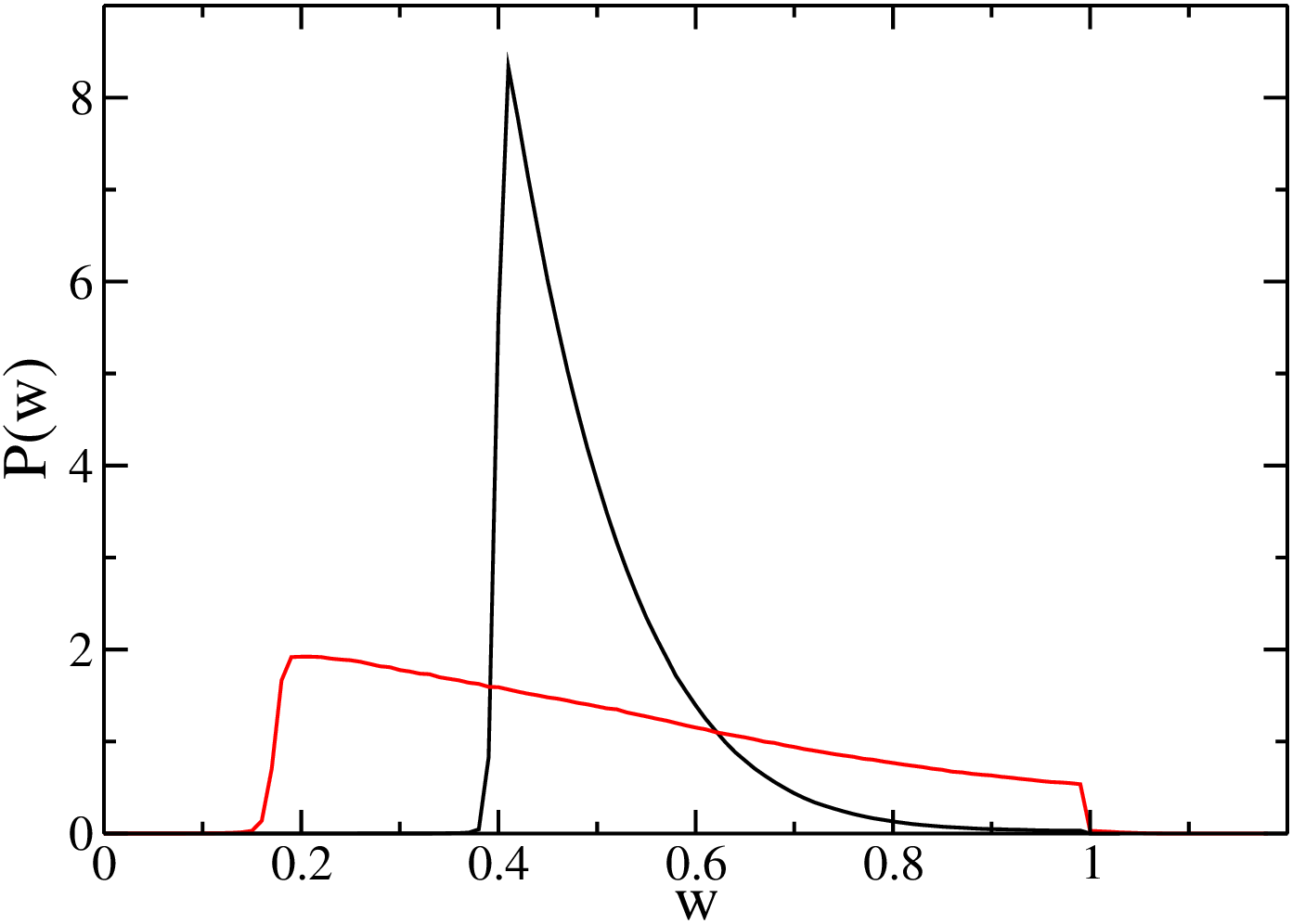}
\caption{(color online) Plot of  the wealth distribution for $N=10000$ agents and averaging over $1000$ samples~\cite{PIAV2003}. The sharp (black) line indicates the situation with nearest neighbors interaction, while the wide (red) line corresponds to the mean field version of the model.} \label{fig:CEMM}
\end{figure}

The dynamics of the system is supported on the idea that some measure should be taken
to modify the situation of the poorest agent. In this context, this action is simulated by
an extreme dynamics~\cite{BakSnep}: at each time step, the poorest agent, i.e., the one with the
minimum wealth, will perform (or be the subject of) some action trying to improve its economic state. Then, at each time step one chooses the agent with minimum wealth (the poorest agent) and substitute it by a new agent (or by the same agent) with a new wealth selected at random in the interval $\{0,1 \}$. Since the outcome of any such measure is uncertain, the minimum suffers a random change in its wealth that we call $\Delta w$~\cite{PIAV2003,SI2004}. As the selected site has the minimum wealth there is a higher chance that it increases its wealth, i.e. that $\Delta w>0$. In the first version of the model it was assumed that whatever wealth is gained (or lost) by the poorest agent it will be at the expenses of its neighbors and that $\Delta w$ will be {\it equally} deducted from (or credited to) its two nearest neighbors on the lattice, making the total wealth constant. Numerical simulations on this model showed that, after a transient, the system arrives at a self-organized critical  state with a stationary wealth distribution that is represented in Fig.~\ref{fig:CEMM}~\cite{PIAV2003}. It is possible to observe that almost all agents are above a certain threshold or poverty line. This poverty line is self-generated\cite{PIAV2003,manna} and not externally imposed~\cite{Ghosh}
Another possibility is to subtract $\Delta w$ from two agents picked at random. This situation has also been considered~\cite{IGPVA2003,SI2004}, it is the {\it annealed} or mean field version of the model, that corresponds to a situation in which the agents with which the exchange takes place are chosen at random and not based on geographical proximity. The distribution is also plotted in Fig.~\ref{fig:CEMM}: it is almost flat and the poverty line is lower than in the case of nearest neighbors interactions.
In the nearest-neighbor version of the model, one founds a minimum wealth or poverty
line that is $\eta_T \approx 40 \%$ of the maximum initial wealth and above this threshold the distribution of the wealth of the agents is an exponential $P(w) \approx \exp(-w^2/2 \sigma^2)$, with $\sigma=22.8$~\cite{IGPVA2003}. The obtained Gini coefficient, $G$~\cite{Gini}, is very low, of the order of $G=0.1$. In the mean-field case the model exhibits a lower threshold, $\eta_T \approx 20 \%$, and, beyond it, also an exponential distribution with a higher value of $\sigma \approx 56.7$~\cite{IGPVA2003} and of the Gini coefficient $G \approx 0.25$.

One interesting characteristic of the dynamics of this model is that the evolution of the system happens in avalanches.
This is so because there is a higher chance that the active site increases its wealth, then their neighbors will more probably lose wealth and became the active site in the next time step~\cite{SI2004}. Examples of these avalanches can be seen in the original Bak-Sneppen article~\cite{BakSnep} and also in ref.~\cite{PIAV2003}. In both cases the distribution of the size of the avalanches follows a power law with the same exponent.

In the present article we would like to discuss a similar model, but considering different networks. We are going to see that some topological effects on the wealth distribution will be observed, particularly a correlation between the wealth and the connectivity either of a given site or of its neighbor. In the next section we present the results for SF and random ER networks.

\section{CEMM on scale free and random networks}
\label{SF}

We performed simulations in complex ER and SF networks~\cite{AB}.
Random ER networks are characterized by a Poisson degree
distribution with $P(k)=e^{- \langle k \rangle} \langle k \rangle^k
/k!$, where $k$ is the degree of a node and $\langle k \rangle$ is the
average degree. In SF networks the degree distribution is given by
$P(k)\sim k^{-\lambda}$, for $k_{\rm min}\leq k \leq k_{\rm max}$,
where $k_{\rm min}$ is the smallest degree, $k_{\rm max}$ is the
degree cutoff and $\lambda$ is the exponent characterizing the
broadness of the distribution: the smaller the exponent of the SF network the higher is its degree of heterogeneity, because as
$\lambda$ decreases the probability of nodes with high degree (hubs) increases. To construct the SF networks we used the Molloy-Reed algorithm ~\cite{Molloy} (or configurational model) with a ``natural'' cutoff for the highest degree $N^{1/(\lambda-1)}$.

We assume that each node of the lattice is an economic agent, and its initial wealth is a random number in the interval $\{ 0,1\}$. Then we apply the minimum dynamics and we wait a number of time steps long enough for the system to arrive to a stationary state. The number of time steps depends on the size of the network. Here we have simulated networks with $N=10^4$ nodes and the number of time steps to arrive to a stationary distribution is $t=10^6$. The obtained wealth distribution is similar to the original CEMM model, but the poverty line and the inequality  depends on the average connectivity of the network $\langle k \rangle$.

\begin{figure}
\centering
\includegraphics[width=8.cm]{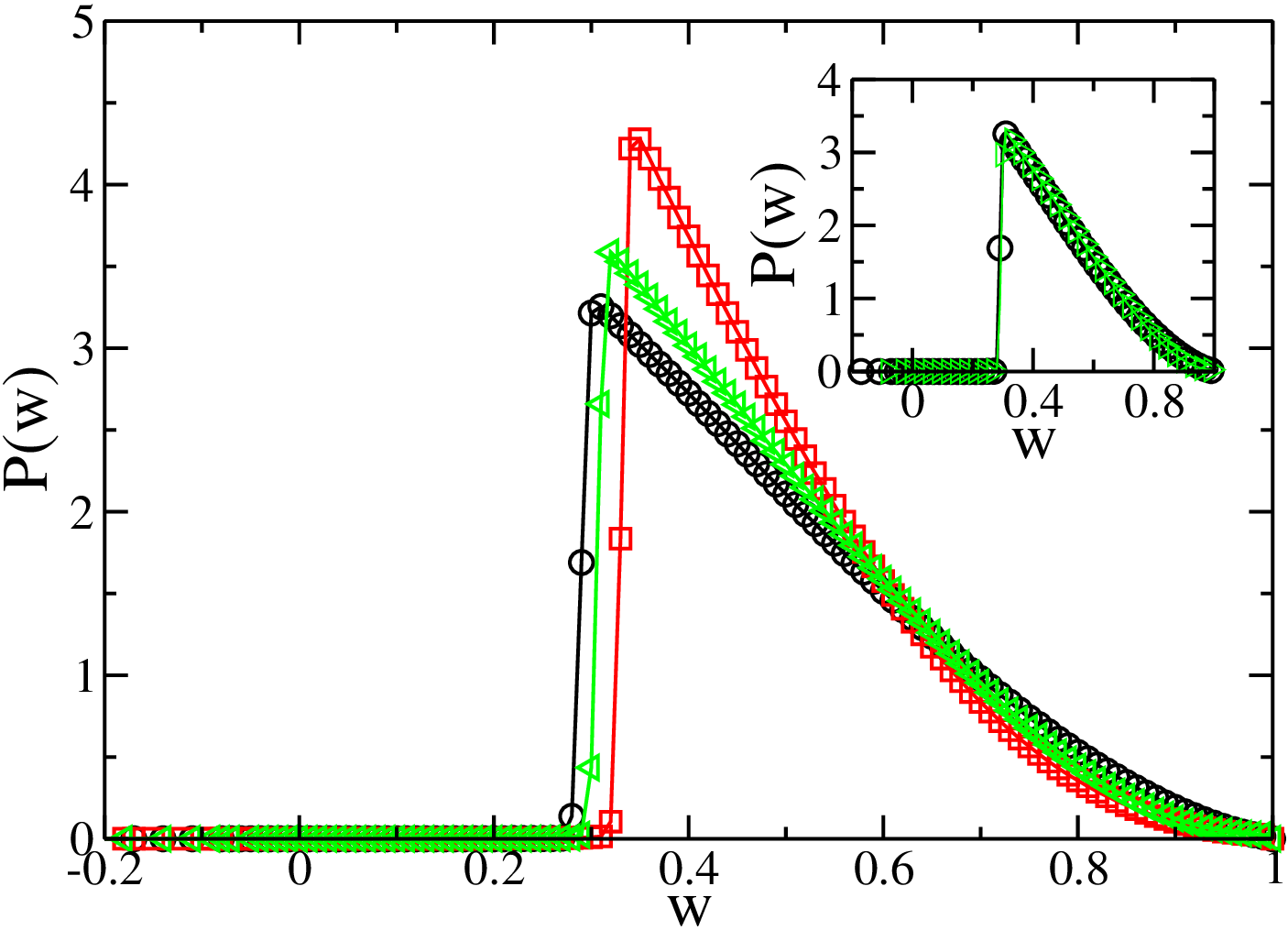}
\caption{(color online) Plot of the wealth distribution for two SF networks with $\lambda=2.5$ (black circles, $\langle k \rangle=5.2$) and $3.5$ (red squares, $\langle k \rangle=2.9$), compared with a random ER network with $\langle k \rangle=4$ (green triangles). In the inset we plot $P(w)$ as a function of $w$ for $\langle k \rangle=5.2$ for both SF and ER networks, in order to show that the poverty line is practically the same for both networks.} \label{Fig:dist}
\end{figure}

On Fig.~\ref{Fig:dist} we have represented the wealth distribution for three different networks: two SF networks and one ER network. It is possible to see that in all cases the poverty line is higher than in the mean field case but lower than for the linear chain (both presented on Fig.~\ref{fig:CEMM}). Considering the degree distribution, the poverty line is lower for a SF network with smaller exponent $\lambda$, corresponding to distributions with higher average connectivity. For the random ER network the poverty line lies in between both plotted SF networks.
In the inset of Fig.~\ref{Fig:dist} we represent $P(w)$ as a function of $w$ for two SF and ER networks with the same average degree. We can see that the curves strongly overlap, suggesting that the distribution mainly depends on the average connectivity, $\langle k \rangle$.
Furthermore, the values of the poverty lines are very close to each other.

\begin{figure}
\centering
\includegraphics[width=8.cm]{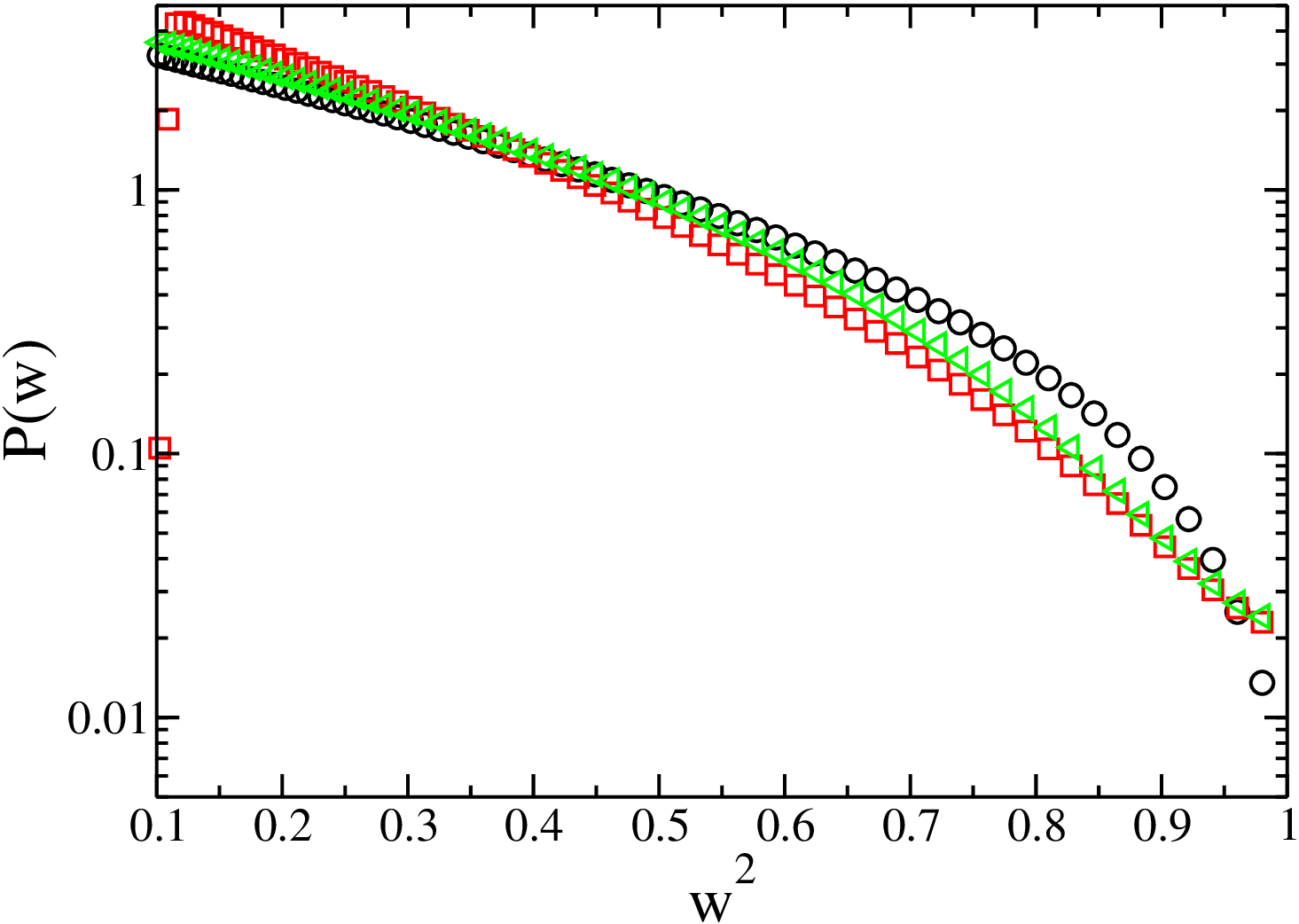}
\caption{(color online) Log-Linear Plot of the fraction of agents with a given wealth as a function of the square of the wealth, for two SF networks with $\lambda=2.5$, $\langle k \rangle=5.2$  (black circles) and $\lambda=3.5$, $\langle k \rangle= 2.9$ (red squares), compared with a random ER network with $\langle k \rangle=4$ (green triangles).\label{Fig:exp2}}
\end{figure}

Also, in order to compare with the original model, we have represented the wealth distribution in a semi-log plot against the square of the wealth. In the original model the number of agents with a given wealth was numerically found to be a truncated exponential function of the square of the wealth. In the present situation we can observe that the situation is different. On Fig.~\ref{Fig:exp2} one can verify that the exponential behavior seems to be valid only for low values of the wealth, while there is a clear deviation from the linear behavior and the distribution goes to zero {\it faster} than an exponential for high values of the wealth. It is also clear that the slope of the distribution decreases when the average connectivity increases.


\begin{figure}
\centering
\includegraphics[width=8.cm]{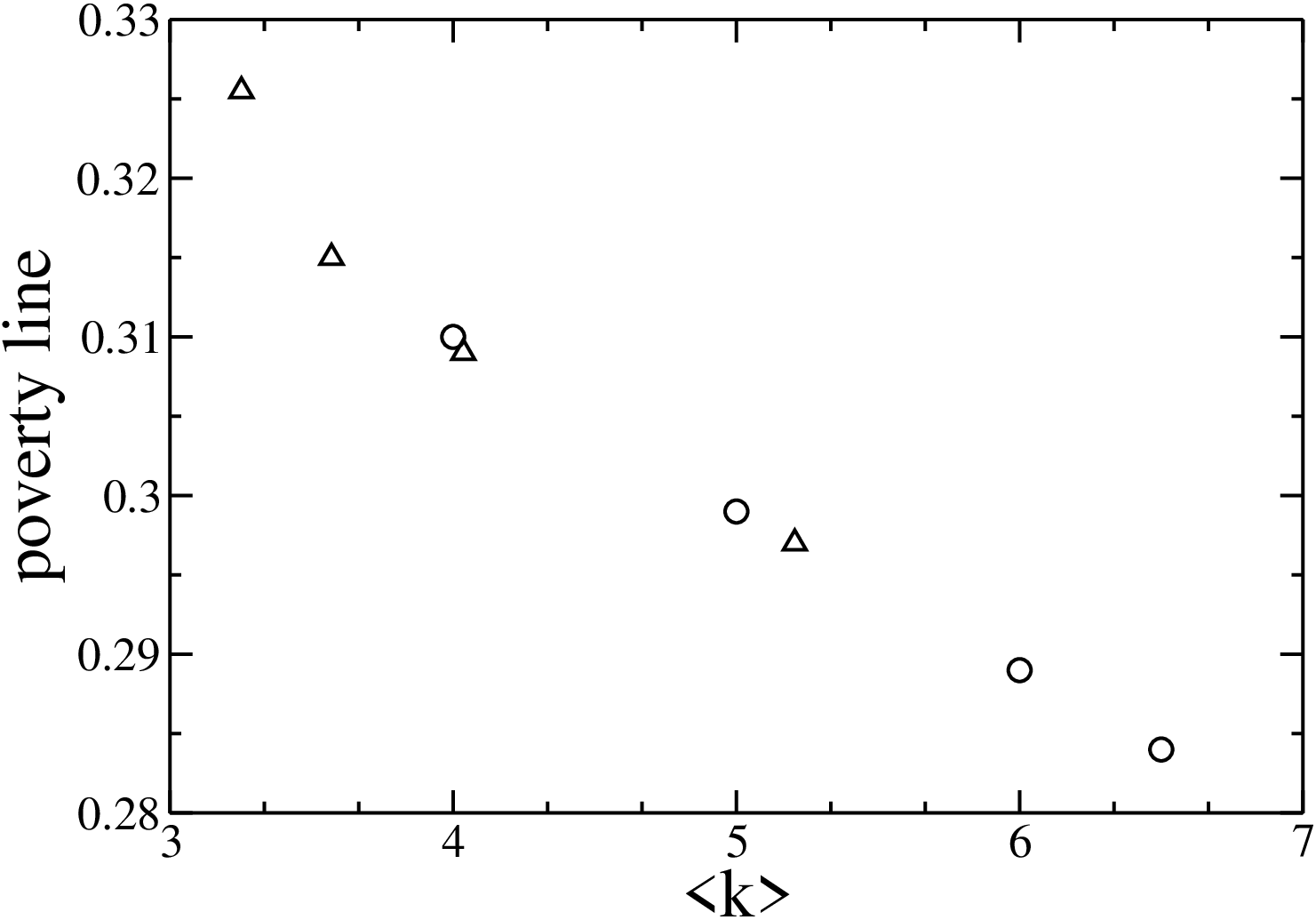}
\caption{Plot of the position of the poverty line as a function of $\langle k \rangle$ for SF networks (circles), where $\langle k \rangle \sim
(\lambda-1)/(\lambda-2)$, and ER networks (triangles). The bigger the average connectivity the lower the poverty line. This is in agreement with the original work, where a low connectivity results in a higher poverty line~\cite{PIAV2003,IGPVA2003}}.  \label{Fig:povline}
\end{figure}

\begin{figure}
\centering
\includegraphics[width=8.cm]{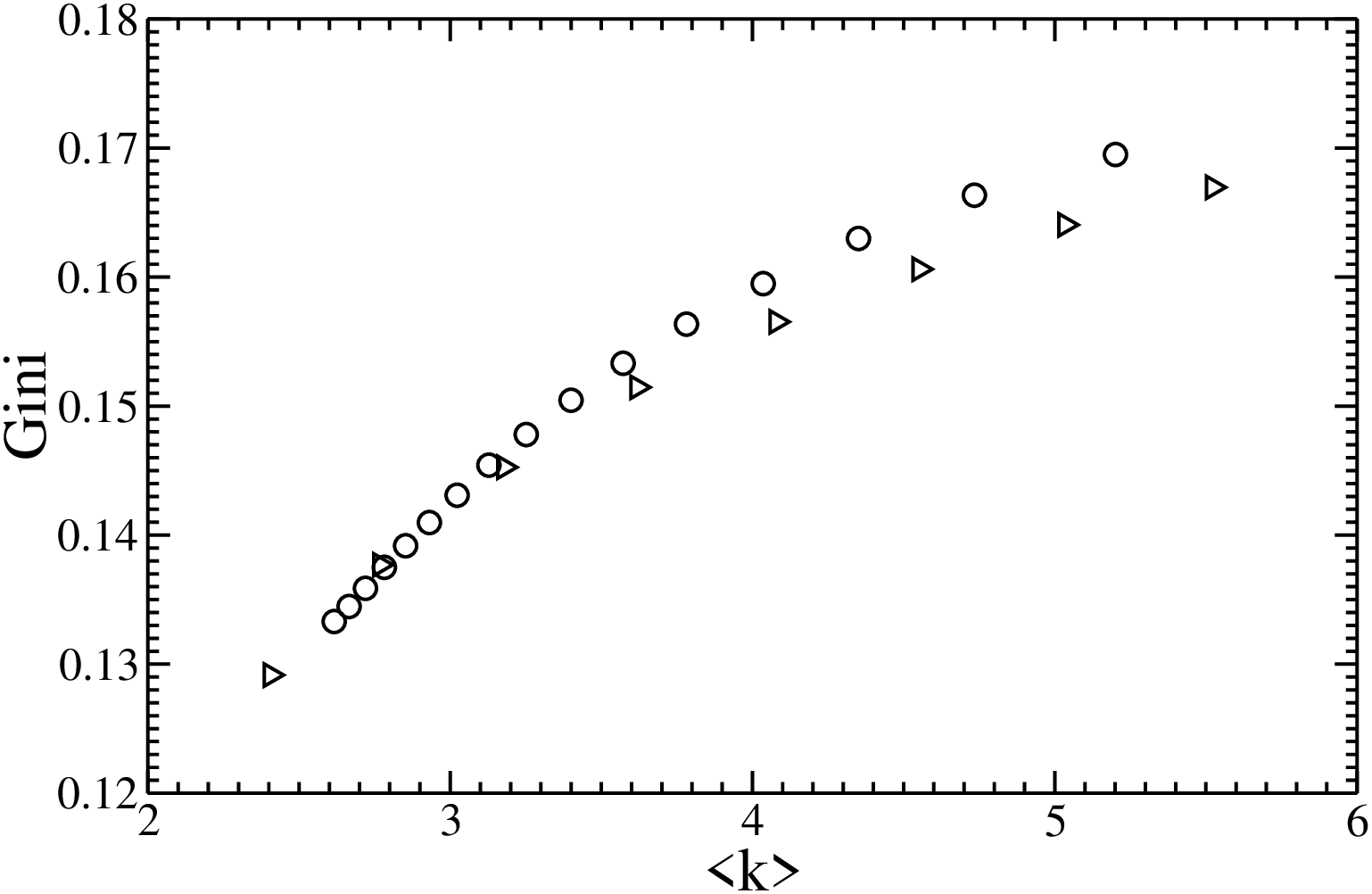}
\caption{Plot of the Gini coefficient as a function of $\langle k \rangle$ for SF networks (circles), where $\langle k \rangle \sim
(\lambda-1)/(\lambda-2)$, and ER networks (triangles). A lower average connectivity diminish the Gini coefficient generating a less unequal society. In any case the values of the Gini coefficient are very low, and compare well with the case of the one-dimensional ring where the obtained Gini coefficient was of the order of $0.08$~\cite{PIAV2003}. We remark that the Gini coefficients are equal for low values of the average degree and present a small difference for higher values of the average connectivity (being the ER the more ``egalitarian'' network, probably because its maximum degree is lower than for an SF network)} \label{Fig:Gini}
\end{figure}

Also, the inset in Fig.~\ref{Fig:dist} suggest that the poverty line is mostly a function of the average connectivity. In order of verify this point we represent on Fig.~\ref{Fig:povline} the position of the poverty line for different SF and ER networks and the results strongly indicate that the poverty line depends only on the average connectivity. Also, the present values may be extrapolated to the case of the linear chain, where the connectivity is $k=2$: the obtained poverty line should be near the value indicated on Fig.\ref{fig:CEMM}, i.e. $\backsim 0.4$. However, it is not just the poverty line that is essentially a function of the average connectivity, there is also a visible effect in the inequality, that we measure through the Gini coefficient. We have represented on Fig.~\ref{Fig:Gini} the Gini coefficient as a function of the average connectivity, $\langle k \rangle$. One can observe that the Gini coefficient decreases when increasing the exponent of the network, i.e. when decreasing the average connectivity. This is so because local interactions produce lower inequality than global interactions~\cite{SI2004}. This is in qualitative agreement with the results of the original CEMM model where a Gini of only $0.08$ was obtained for the one dimensional ring with nearest neighbors interactions ($k=2$)~\cite{PIAV2003}. Nevertheless, the Gini coefficient is slightly smaller for ER networks compared with SF networks with the same average connectivity, probably because the maximum degree of the ER network is lower than for the SF network.

One also should expect that the degree of the agent should be relevant to determine the chances of being affected by its neighbors. We discussed in the previous section that the evolution of the system occurs through avalanches. Then the neighbors of the active site have a higher probability of becoming active in the following time step. If it is the case sites with more neighbors (high degree) will have more chances of becoming active or, in other words, to be the new poorer one. Then, it should be more convenient to have few neighbors than to be highly connected.

\begin{figure}
\includegraphics[width=7cm]{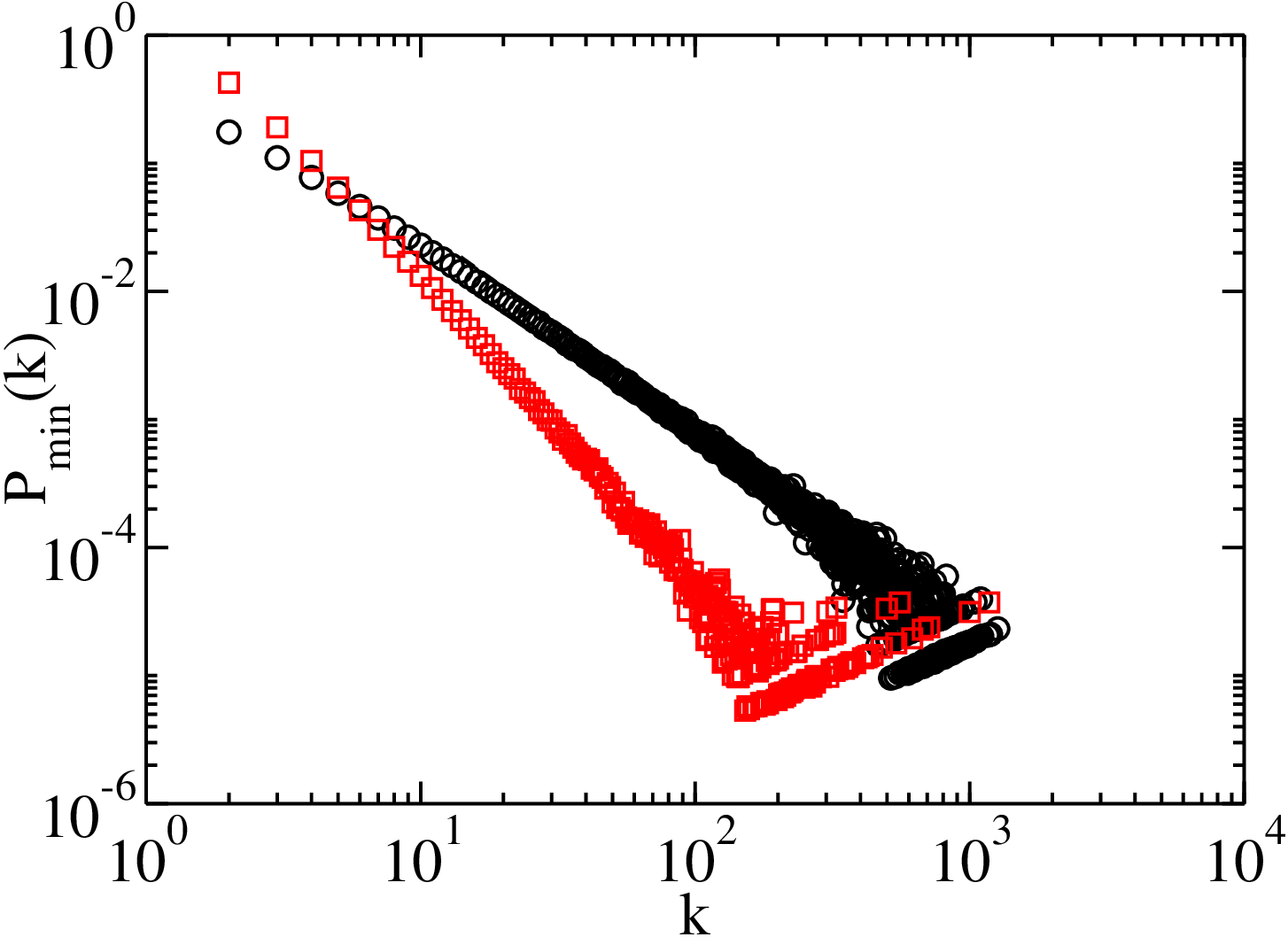}
\includegraphics[width=7cm]{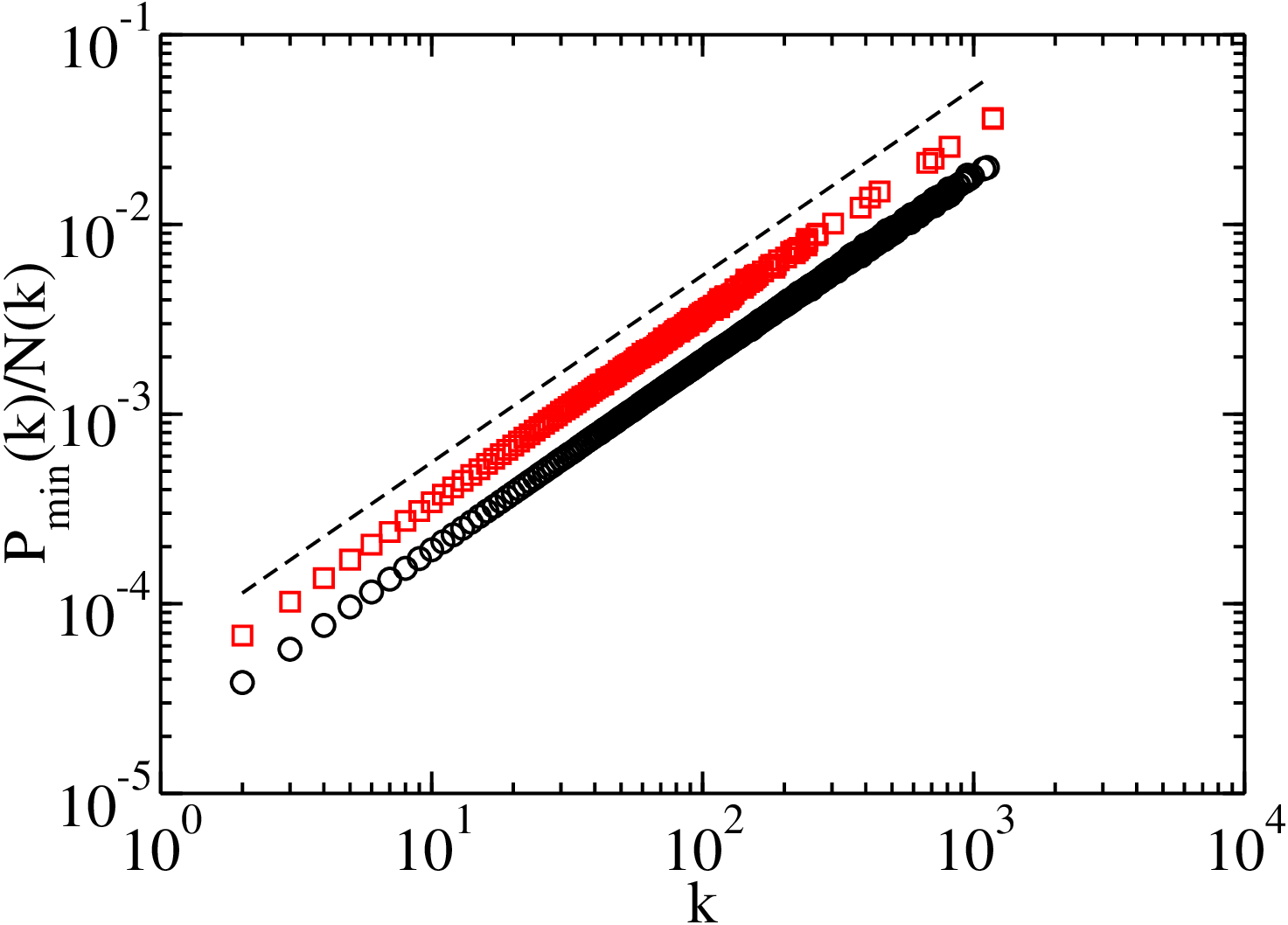}
\caption{(color online) Left panel: Plot of the number of times any site of connectivity $k$ is the site with minimum wealth (the poorest one) as a function of its degree. Right panel: Plot of the frequency any site of degree $k$ is the poorest one divided by the number of sites of degree $k$ and as a function of the degree. Black circles correspond to a network with exponent $\lambda=2.5$ and red squares to $\lambda=3.5$. The dashed line in the right panel indicates the slope $1$. The right panel confirms that high degree nodes are the minimum wealth ones with higher frequency} \label{Fig:degree_corr}
\end{figure}

We can verify this result by looking at figure~\ref{Fig:degree_corr} where we have represented the number of times any site of connectivity $k$ is the site with minimum wealth as a function of its grade: The left panel seems to indicate the opposite of what we are stating, as the number of times a site is the site of minimum wealth is maximum for $k=1$ and decreases following a power law  with slope $3/2$ for $\lambda = 2.5$ and a slope a little bit higher for $\lambda = 3.5$ But we must be aware that in a SF network the degree distribution is a power law, there are much more nodes with low degree than with high degree. If we want to determine the frequency that any node with a given degree is the poorest one, we must divide the number of times any node with degree $k$ is the minimum by the number of nodes with degree $k$: $N(k)$. This result is plotted in the right panel of  figure~\ref{Fig:degree_corr} and confirm our prediction. Moreover, the frequency at which a node of degree $k$ is the minimum increases linearly with $k$ and the slope is almost $1$ {\it independently of the exponent $\lambda$ of the network}.

\begin{figure}
\includegraphics[width=7.cm]{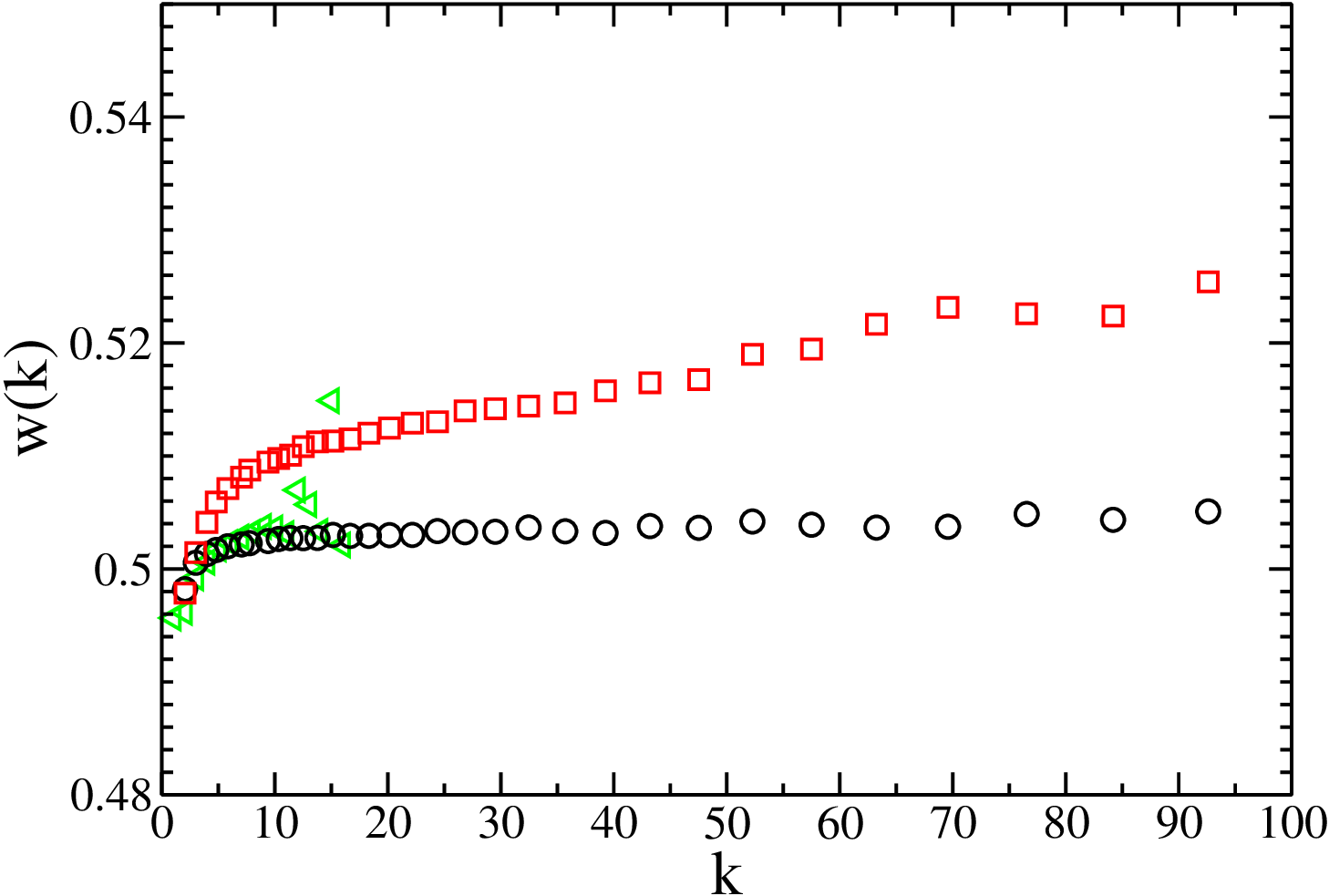}
\includegraphics[width=7.cm]{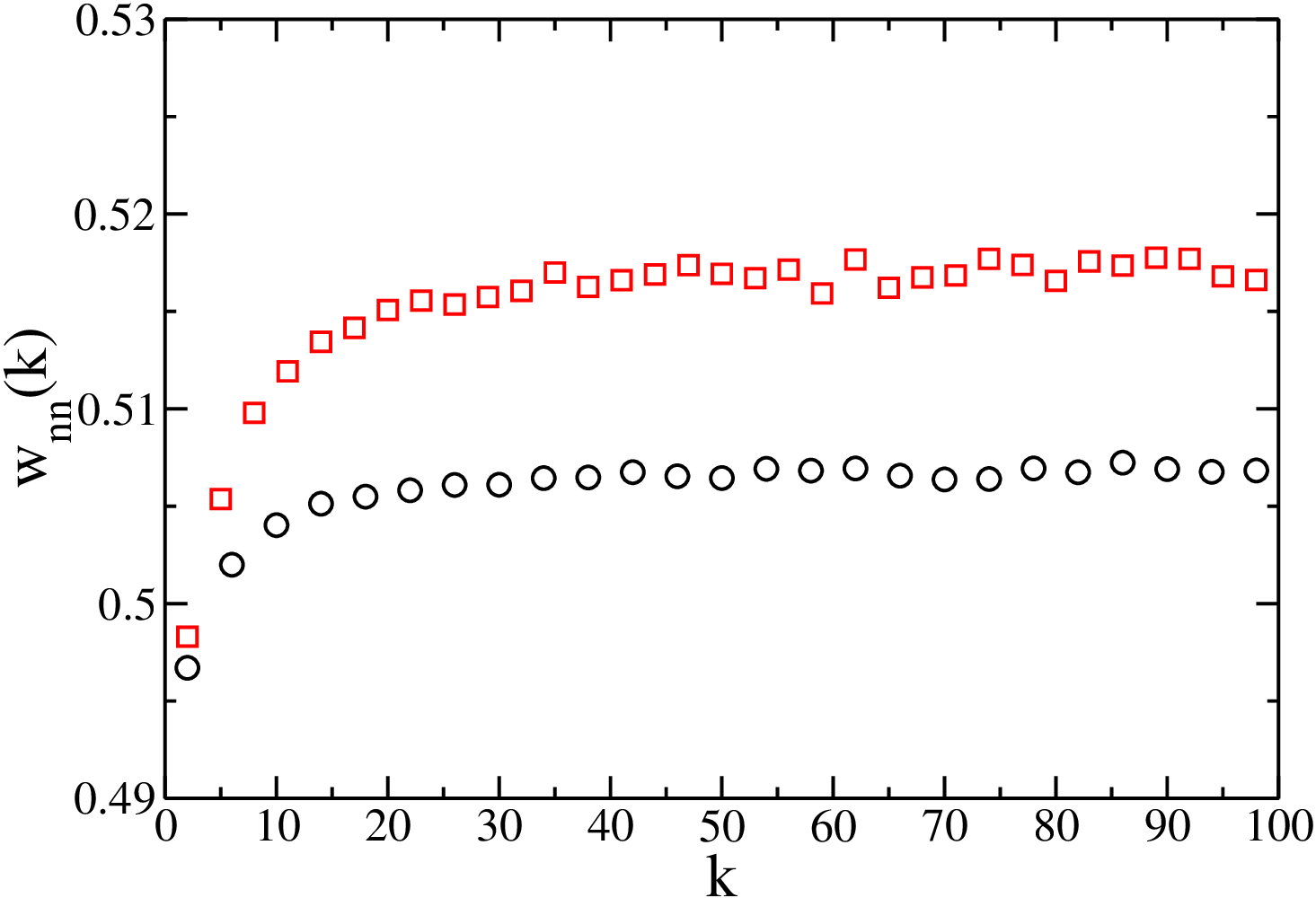}
\caption{(color online) Left panel: Plot of the average wealth of a node of grade $k$, $w(k)$. Right panel: Plot of the average wealth of the neighbors of a node with degree $k$, $w_{nn}(k)$ . Both panels represent a SF network with $\lambda=2.5$ (black circles) $ \lambda=3.5$ (red squares). Green triangles represent the ER network with $\langle k \rangle=4$ (it is not possible to identify neighbors in the ER network, so there is no representation of the ER network in the right panel). Remark that the neighbors of low connectivity nodes are in disgrace, as their average wealth is lower than the global one: $0.5$.} \label{Fig:knn}
\end{figure}

As a verification of the previous result we have calculated the average wealth as a function of the number of links an agent has (the global average wealth is constant, as the system is conservative). We define de average wealth of a node of degree $k$ as the sum of the wealth of the nodes with degree equal to $k$ divided by the number of nodes with degree $k$, $N(k)$ i.e. $w(k)= \frac{\sum_{i \in \Omega_k} w_i}{N(k)}$ where $i=1,...N$ and $\Omega_k$ is the set of nodes with degree $k$.
We present in the left panel of figure~\ref{Fig:knn} this average wealth as a function of the degree: once again the results seems to be in contradiction with the conclusion that it is more convenient to have low degree, as the figure shows that low degree nodes have, in average, a wealth lower than the global average wealth. But one should keep in mind that the number of low degree nodes is much bigger than the rest, so it is reasonable to think that they possess a huge fraction of the total wealth but, in average, the wealth {\it per capita} is somewhat smaller than the global average. On the other hand the average wealth of the high degree nodes is slightly higher than the global average. That means that high degree nodes are minimum with higher frequency but, in average, they have a wealth higher than the global average. However it is even better to be the neighbor of a high degree node. In the right panel of figure~\ref{Fig:knn} we plot the average wealth of the neighbors of a node with degree $k$, defined as $w_{nn}(k)= \frac{\sum_{i \in \Omega_k} \sum_{j \in \Lambda_i} w_j}{k N(k)}$ where $\Lambda_i$ is the set of the $k$ neighbors of node $i$. In this last figure (right panel of Fig.\ref{Fig:knn}) it is clear that the neighbors of low $k$ nodes are less favored while neighbors of nodes with $k \geq 5$ exhibit an average wealth higher than the global average. The figures correspond to networks with $\lambda = 2.5$ and $\lambda=3.5$ but the results are qualitatively the same for networks with different values of $\lambda$. Thus, the last sentence of the previous paragraph should be modified by saying that in order {\it to have a wealth higher than the average one must have high degree and/or be the neighbor of one high degree node}.

\section{Conclusions}
\label{conclu}

In the present article we have studied the Conservative Exchanges Market Model (CEMM) on different types of complex networks. We have verified that some characteristics of the models are robust, as for example the existence of a poverty line for finite wealth, and the fact that disorder on the lattice increases the inequality (i.e. the Gini coefficient) and moves the poverty line to lower values of income. On the other hand, deviations from the truncated exponential dependence of the wealth distribution on the square of the wealth, observed in the original model~\cite{PIAV2003,IGPVA2003} and in a small world network~\cite{IGVA2004} are much more evident in SF or ER networks, as it was shown on Fig.~\ref{Fig:exp2}.

The more interesting effects are related to the correlation of the wealth with the connectivity of the agents. These correlations are very important in SF networks. One expected result, verified by our calculations, is that in a SF network, highly connected nodes, ``hubs'', are richer than the average. But this is true when making an instantaneous picture of the distribution of wealth on the network (see Fig.~\ref{Fig:knn}). The analysis of the time evolution indicates that the life of these richer ``hubs'' is not easy: they become the minimum wealth agents with a frequency higher than low connected agents (see Fig.~\ref{Fig:degree_corr}). This is so because they have lots of lower degree neighbors and, as one can verify looking to the right panel of Fig.~\ref{Fig:degree_corr}, agents that are neighbors of low degree nodes have a wealth lower than the average.

Thus, SF networks provide a good description of human societies. High connected agents (people, companies, banks) are richer in average but subjected to strong fluctuations, while low connected agents are poorer. In any case this model provides a poverty line that guarantees a relatively low Gini coefficient, thus the state of the poorer agents is not as catastrophic as in other exchange models~\cite{Caon2007,rita2012}: within the CEMM's there is no condensation of wealth. A drawback of the model could be the fact that the networks are static, a better representation of social economic exchanges should include the possibility of changing partners, i.e. evolving social networks. Work in this direction is in progress.

\section*{Acknowledgements}
We acknowledge fruitful discussions with G. Abramson, S. Gon\c{c}alves, F. Laguna, R. Don\'angelo and J.L. Vega. JRI acknowledges the kind hospitality of the Departamento de F\' {\i}sica of Universidad Nacional de Mar del Plata where this work was started, and financial support from Brazilian agencies CNPq, FAPERGS and CAPES. LAB and PAM thanks the financial support of UNMdP and FONCyT: Pict 0293/08.



\begin{thebibliography}{99}

\bibitem{Pareto} Pareto V (1897), Cours d'Economie Politique, Vol. 2, F. Pichou, Lausanne, Switzerland

\bibitem{Mantegna} Mantegna R M and Stanley H E (2000), An Introduction to Econophysics, correlations and complexity in Finances, Cambridge University Press, Cambridge, UK

\bibitem{souma} Aoyama H Souma W and Fujiwara Y (2003) Physica A 324:352

\bibitem{clementi} Clementi F and Gallegati M (2005) Physica A 350:427

\bibitem{sinha1} Sinha S (2006) Physica A 359:555

\bibitem{dragu2000} Dragulescu A and Yakovenko V M (2000) The European J. of Physics B
17:723

\bibitem{dragu2001a} Dragulescu A and Yakovenko V M (2001) The European J. of Physics B 20:585

\bibitem{dragu2001b} Dragulescu A and Yakovenko V M (2001) Physica A 299:213

\bibitem{PerBak} Bak P (1999), How Nature Works, Springer-Verlag, New York

\bibitem{Caon2007} Caon G M , Gon\c{c}alves S and Iglesias J R, (2007) The European Physical Journal - Special Topics,
143:69

\bibitem{PIAV2003} Pianegonda S, Iglesias J R, Abramson G and Vega J L (2003) Physica A 322:667

\bibitem{SI2004} Pianegonda S and Iglesias J R (2004) Physica A 342:193

\bibitem{sinha2} Sinha S (2003)  Physica Scripta T106:59

\bibitem{chatter1} Chatterjee A, Chakrabarti B K and Manna S S (2004) Physica A 335:155

\bibitem{chatter2}Chakrabarti B K and Chatterjee A (2004), Ideal Gas-Like Distributions in Economics: Effects of Saving Propensity, in ``Applications of Econophysics'', Ed. H. Takayasu, Conference proceedings of Second Nikkei Symposium
on Econophysics, Tokyo, Japan, 2002, by Springer-Verlag, Tokyo.
Pages 280--285. Preprint available: cond-mat/0302147.

\bibitem{chakra} Chakraborti A and Charkrabarti B K (2000) The European J. of
Physics B 17:167

\bibitem{IGPVA2003} Iglesias J R, Gon\c{c}alves S, Pianegonda S, Vega J L and Abramson G (2003)
Physica A 327:12

\bibitem{IGVA2004} Iglesias J R, Gon\c{c}alves S, Abramson G and Vega J L (2004)
Physica A 342:186

\bibitem{Bouchaud} Bouchaud J.-P. and M\'ezard M., (2000) Physica A 282:536

\bibitem{cristian07} Mourkazel C F, Gon\c{c}alves S, Iglesias J R,
Achach M, and Huerta R (2007) The European Physical Journal - Special Topics,
143:75

\bibitem{rita2012} Iglesias J R and de Almeida R M C, (2012) Eur. Physical J. B 85:85

\bibitem{west} Scafetta N, West B J and Picozzi S (2003) cond-mat/0209373v1(2002) and cond-mat/0306579v2.

\bibitem{hayes} Hayes B, (2002) American Scientist, 90:400

\bibitem{Rawls} Rawls J (1971), A Theory of Justice, The Belknap Press of Harvard University Press, Cambridge, MA.

\bibitem{Sen} Sen A (2009) The Idea of Justice, The Belknap Press of Harvard University Press, Cambridge, MA.

\bibitem{BakSnep} Bak P and Sneppen K (1993) Phys. Rev. Lett. 71:4083

\bibitem{manna} Chakraborty Abhijit, Mukherjee G and Manna S S (2012), Fractals 20:163

\bibitem{Igles2010} Iglesias J R, (2010) Science and Culture (India) 76:437

\bibitem{Ghosh} Ghosh A, Basu U, Chakraborti A and Chakrabarti B K, (2011) Phys. Rev. E 061130

\bibitem{Gini} http://en.wikipedia.org/wiki/Gini\_coefficient

\bibitem{AB} Albert R and Barab\'asi A-L (2002), Rev. Mod. Phys. 74:47

\bibitem{Molloy} M. Molloy and B. Reed, (1998) Comb. Probab. Comput. 7:295;
M. Molloy and B. Reed, (1995) Random Struct. Algorithms 6:161.

\end{thebibliography}
\end{document}